\documentclass{jkas}

\def\year{2014} 
\def\month{May} 
\def\volume{00} 
\def\issue{0} 
\def\beginpage{1} 
\def\endpage{99} 
\def\received{February 30, 2014} 
\def\accepted{February 31, 2014} 
\date{Received \received; accepted \accepted}

\newcommand\ion[2]{{#1}\,{\sc #2}} 

\title{
A New Method to Determine the Temperature of CMEs Using a Coronagraph Filter System
}

\author[1]{Kyuhyoun Cho}
\author[1]{Jongchul Chae}
\author[2]{Eun-kyung Lim}
\author[2, 3]{Kyung-suk Cho}
\author[2, 3]{Su-Chan Bong}
\author[1]{Heesu Yang}

\affil[1]{Department of Physics and Astronomy, Seoul National University, Gwanak-gu, Seoul 151742, Korea \email{chokh@astro.snu.ac.kr}}
\affil[2]{Korea Astronomy and Space Science Institute, 776 Daedeokdae-ro, Yuseong-gu, Daejeon 305348, Korea}
\affil[3]{University of Science and Technology, 217 Gajeong-ro, Yuseong-gu, Daejeon 305350, Korea}

\def\corrauthor{
K. Cho
}

\def\runningauthor{
Cho et al.
}

\def\runningtitle{
Temperature Determination of CMEs Using a Coronagraph Filter System
}

\def\keywords{
Sun: coronal mass ejections (CMEs) --- method: numerical
}

\def\abstracttext{
The coronagraph is an instrument enables the investigation of faint features in the vicinity of the Sun, particularly coronal mass ejections (CMEs). So far coronagraphic observations have been mainly used to determine the geometric and kinematic parameters of CMEs. Here, we introduce a new method for the determination of CME temperature using a two filter (4025 \AA~and 3934 \AA) coronagraph system. The thermal motion of free electrons in CMEs broadens the absorption lines in the optical spectra that are produced by the Thomson scattering of visible light originating in the photosphere, which affects the intensity ratio at two different wavelengths. Thus the CME temperature can be inferred from the intensity ratio measured by the two filter coronagraph system. We demonstrate the method by invoking the graduated cylindrical shell (GCS) model for the 3 dimensional CME density distribution and  discuss its significance.
}

\begin{document}
\jkashead 


\section{Introduction\label{sec:intro}}

 The coronagraph is an instrument to observe the faint structure outside the solar disk. By screening intense radiation from the solar disk, it enables us to observe the K-corona, streamers, coronal mass ejections (CMEs), and other phenomena in the corona. The origin of visible light detected by coronagraphic observations is Thomson scattered light from the solar photosphere by free electrons in the solar corona; its properties depend on the electron density distribution, the polarization, and the scattering angle.

 For decades, our knowledge about CMEs has greatly increased through white light coronagraphic observations. A CME expels huge amounts of plasma from the solar atmosphere and is one of the most energetic events in the heliosphere. Since \citet{Tousey1973} carried out the first CME observation using the OSO-7 coronagraph, succeeding spacecraft coronagraphic observations allowed the investigation of CMEs. Now, it is generally accepted that  CMEs have three-part structures \citep{Illing1985} and are responsible for changes in space weather that affect interplanetary space and terrestrial magnetism \citep{Baker2008}. A number of statistical studies on angular widths \citep{Howard1985, StCyr2000}, velocities \citep{Yashiro2004}, and occurrence latitudes \citep{Hundhausen1993, Gopalswamy2010} were performed with accumulated coronagraphic observation data. But the derived characteristics of CMEs from coronagraphic observations were limited to mainly geometric and kinematic parameters.

 Observational reports of CME temperatures are scarce due to the difficulty of spectroscopic observations of CMEs, which are required for temperature determination. The CME temperature can be inferred either from temperature-sensitive line ratios or from the existence of emission lines of  specific formation temperatures. Although the spectral profiles in principle can give us the temperature directly, they do not provide spatial information.  One can obtain useful spectra only when the slit position is cospatial with the CMEs. Only a few studies reported CME temperatures using the Ultraviolet Coronagraph Spectrometer (UVCS) on board the Solar and Heliospheric Observatory (SOHO) \citep{Ciaravella2000, Akmal2001}.

 Recently, the technique of  differential emission measure (DEM) was used to estimate the CME temperature \citep{Lee2009, Hannah2013}. The Atmospheric Imaging Assembly (AIA) on board the Solar Dynamics Observatory (SDO) or the EUV Imaging Spectrometer (EIS) on board the Hinode can observe CMEs lower in the solar atmosphere at various EUV wavelengths. The different temperature dependence of  the EUV filters allows us to estimate the multi-thermal structure of CMEs with the DEM method. This method requires the combination of many EUV filters and has height limitations because the EUV filters are usually dedicated to the solar disk observations.

\begin{figure*}[!t]
\centering
\includegraphics[angle=0,width=150mm]{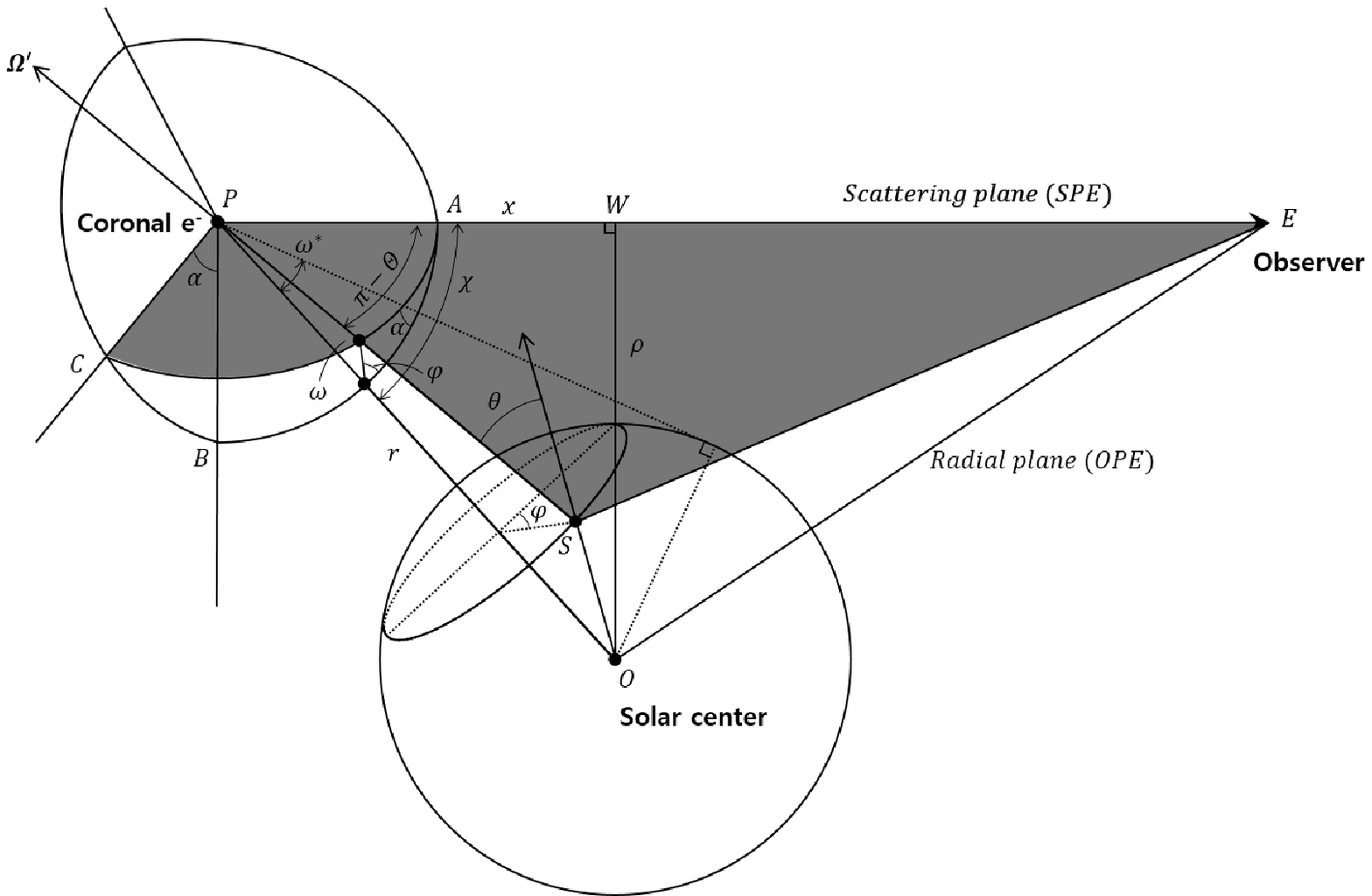}
\caption{Geometry for the K-coronal spectrum calculation from \citet{Reginald2001}.\label{fig:jkasfig1}}
\vspace{5mm} 
\end{figure*}

 We introduce a new method for the determination of CME temperature through coronagraphic observations. Our method is very simple and has no spatial limitations compared to previous methods. \citet{Reginald2009} measured the electron temperature of the inner K-corona using filter observation during the total solar eclipse. We extend this method to the CME temperature measurement using coronagraph filter observations. We consider an isothermal CME with a density structure that is described by a gradual cylindrical shell (GCS) model  and calculate the expected filter intensity ratio maps for 4 different CME temperatures using this simple method. We present our results and discuss the importance and the prospects of the method.

\section{Method and Result\label{sec:result}}

 \citet{Cram1976} argued the possibility of coronal temperature measurement from the specific intensity ratio. He noted that strong absorption lines in the visible light irradiated from the solar photosphere are flattened by the Doppler effect from thermal motions of free coronal electrons. High electron temperature flattens the spectrum, and, hence,  the intensity ratio between the absorption line center and the continuum decreases. 
 Thus the intensity ratio can be used as an indicator for electron temperature. Since stronger absorption lines are more favorable for measuring the intensity ratio difference,  the filters are usually put  around  4000 \AA~where the strongest lines of the \ion{Ca}{ii} H \& K and the G band exist.

 \citet{Reginald2001} improved the accuracy of the method by adding the solar wind effect. He supposed that coronal electrons are moving in the radial direction with a constant solar wind speed. In this situation, the light from the solar photosphere  appears redshifted to the coronal electrons that are receding from the photosphere.  This redshifted light is scattered by the coronal electrons to the observer. If the scattering electron is positioned behind the solar limb plane, the light is further shifted to the red, enhancing the redshift. On the other hand, if the electron is positioned in front of the solar limb plane, the light is shifted back reducing the redshift (see Figure 3 in \citealt{Reginald2009}). Thus, the total effect of the solar wind on the scattered light is redshift. It means that a faster solar wind shifts the K-coronal spectrum to longer wavelength and affects the intensity ratio more strongly.
\begin{figure*}[!t]
\centering
\includegraphics[angle=0,width=150mm]{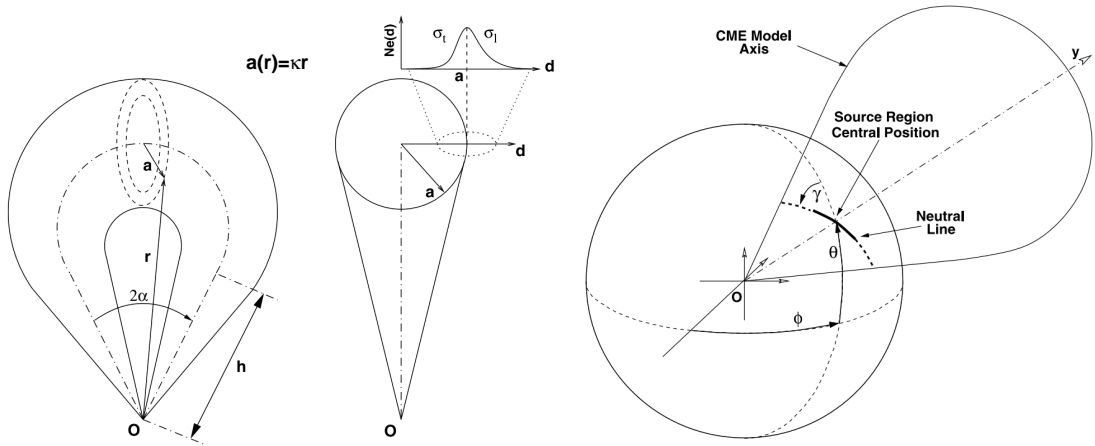}
\caption{Face-on, edge-on, and 3D representation of the GCS model from \citet{Thernisien2006}. \label{fig:jkasfig2}}
\vspace{5mm} 
\end{figure*}
\begin{table*}[t]
\caption{Model and Positioning Parameters of the GCS model for 2002 January 4 CME from \citet{Thernisien2006}.\label{tab:jkastable1}}
\centering
\begin{tabular}{ccccccccc}
\toprule
$\alpha$ (deg) & $\kappa$ & $h$ $(R_{\odot})$ & $n_{e}$ $(cm^{-3})$ & $\sigma_{t}$ & $\sigma_{l}$ & $\varphi$ (deg) & $\theta$ (deg) & $\gamma$ (deg)
 \\
\midrule

26.9 & 0.43 & 1.48 & 8.69 $\times 10^{5}$ & 0.2 & 0.28 & 326 & 25 & 62 \\
\bottomrule
\end{tabular}
\end{table*}

 Figure \ref{fig:jkasfig1} shows the geometry for K coronal spectrum calculation used by \citet{Reginald2001}. The light from the solar photosphere (S) is scattered by the free electrons in the corona (P) to the observer (E). For this geometry, \citet{Reginald2001} used the following calculation formula.

\begin{eqnarray}
\mathcal{I}^{S}_{\lambda}(\rho) &=& \int^{\infty}_{-\infty} \mathrm{d} x N_{e}(x) \times \nonumber \\
&~& \int^{2\pi}_{0}\mathrm{d} \varphi \int^{\omega^{*}}_{0} \mathrm{d} \omega \sin{\omega}\, Q^{S}(\omega, \varphi) \times \nonumber \\
&~& \int^{\infty}_{-\infty}   \mathrm{d} \lambda' \;
 I_{\lambda'}(\omega, \varphi)\frac{1}{2\sqrt{\pi}\Delta b} \times \nonumber \\
&~& \qquad e^{-\left[ \left\{\lambda-\lambda'(1+2b^{2}w\cos\omega/c)\right\}/(2\Delta b) \right]^{2}} . \nonumber
 \label{4d int}
\end{eqnarray}

  This equation  consists of 4 dimensional integrations. First, the free electrons on each segment ($\mathrm{d}x$) of  the line of sight (LOS) contribute to the observable intensity. We should consider the LOS distribution of the free electrons, $N_{e}(x)$. \citet{Reginald2001} applied the Baumbach model \citep{Baumbach1937} for the distribution of the K-corona free electrons on the LOS. Second, the  light is integrated over the solar disk or over the solid angle ($\mathrm{d} \varphi \, \mathrm{d} \omega $) of the solar surface spanned by the free electrons (P) in the corona. Limb darkening and polarization are taken into account in this step. Lastly, the wavelength of the incident light ($I_{\lambda'}$) is changed because of the Doppler shift caused by free electron's motion. \citet{Reginald2001} assumed that the main motions for the coronal free electrons are thermal motion of constant K-coronal  temperature and constant radial solar wind speed. Thus, the contribution of the Doppler shift of all the motions should be integrated over wavelength ($\mathrm{d} \lambda'$). By applying this method, \citet{Reginald2009} obtained the K-coronal temperature map from the total eclipse observation.

 The  CME spectrum calculation can be conducted using the same method that was used for the K-coronal intensity. This is possible because the  emission mechanism is the same for both the K corona and CMEs: Thomson scattering. The geometry, however, is considerably different. A CME is not spherically symmetric as the K-corona.  We adopt the GCS model for the 3 dimensional CME density distribution (See Figure \ref{fig:jkasfig2}). \citet{Thernisien2006} tested the GCS modeling technique on 34 LASCO CMEs. Among them, we choose the 2002 January 4 event and adopt their detail parameters for this case study  \citep{Thernisien2010}. The adopted parameters are summarized in Table~\ref{tab:jkastable1}. We use a CME speed of 896 km s$^{-1}$ which is recorded in the SOHO/LASCO CME catalog\footnote{\url{http://cdaw.gsfc.nasa.gov/CME_list}}. The limb darkening coefficients are extracted from \citet{Allen1973}. The extraterrestrial solar spectral irradiance for the incident light from the solar photosphere is provided by \citet{Kurucz2005}\footnote{\url{http://Kurucz.harvard.edu/sun/irradiance2005/irradthu.dat}}. We calculate the spectra in the vicinity region of the CME with CME temperatures of 0.5, 1.0, 1.5, and 2.0 $\times 10^{6}$ K.

 From the simulated spectra, we calculate the two filter intensity ratio for every calculated position. To determine the center wavelength of the filters, we simulate the spherically symmetric coronal spectrum and calculate the wavelengths at which the filter intensity ratio change is maximized with respect to the temperature change. From the result, we select the Gaussian filters which have a width of 30 \AA~and center of 4025 \AA~and 3934 \AA~, respectively. These wavelengths are slightly different from 3850 \AA~and 4100 \AA~used by \citet{Reginald2009}, possibly due to differences in the spectrum simulation or wavelength selection method. Finally we obtain a filter intensity ratio between the two filters (I(4025 \AA~)/I(3934 \AA~)) for every spatial pixel near the CME.

\begin{figure*}[!t]
\centering
\includegraphics[angle=0,width=140mm]{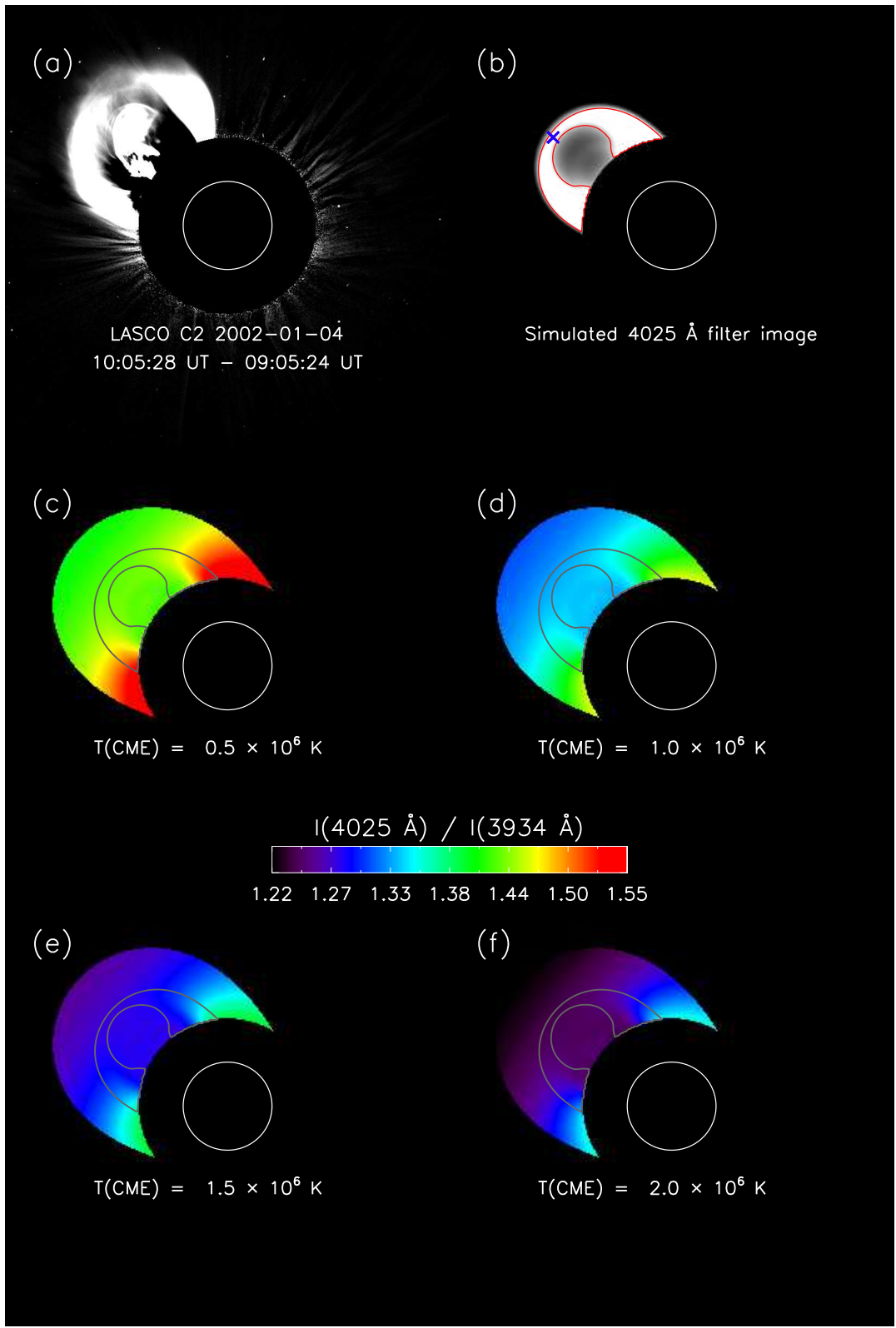}
\vspace{1cm}
\caption{(a) LASCO C2 difference image. (b) simulated 4025 \AA~filter image. (c)-(f) Calculated two filter intensity ratio images with CME temperature of 0.5, 1.0, 1.5, and 2.0 $\times~10^{6}$ K. The gray contours represent the simulated CME structure in Figure \ref{fig:jkasfig3}b.\label{fig:jkasfig3}}
\end{figure*}

\begin{figure*}[!t]
\centering
\includegraphics[angle=0,width=150mm]{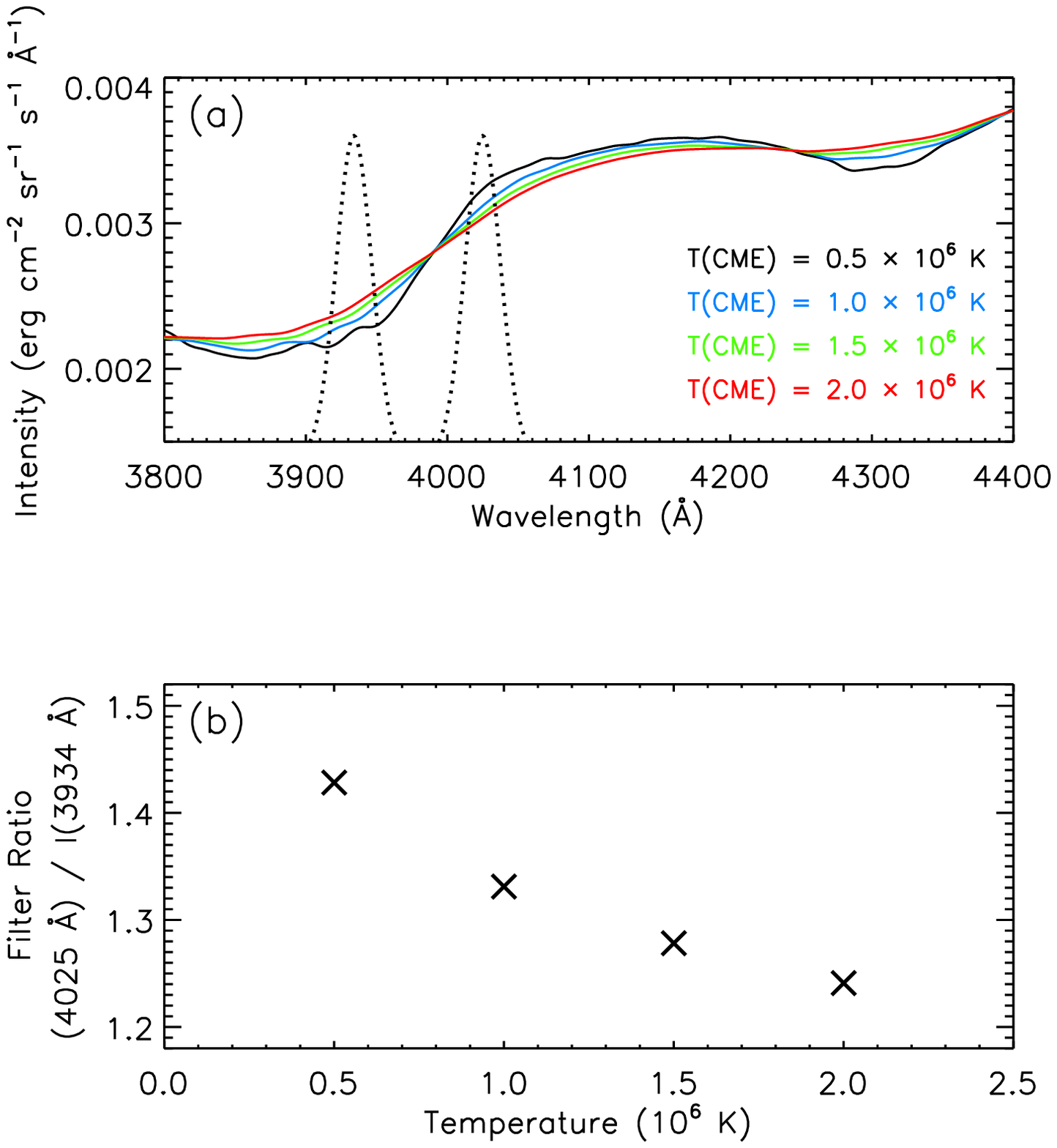}
\caption{(a) CME spectrum at the blue cross position in Figure \ref{fig:jkasfig3}b with CME temperature of 0.5, 1.0, 1.5, and 2.0 $\times~10^{6}$ K. The dashed Gaussian profiles represent filter response functions with centers of 3934 \AA~and 4025 \AA, respectively. (b) Relation between CME temperature and the filter intensity ratio.\label{fig:jkasfig4}}
\vspace{5mm} 
\end{figure*}

 Figure \ref{fig:jkasfig3} shows the observed CME image using LASCO C2 (a), the simulated CME image at 4025 \AA~(b), and the calculated filter intensity ratio maps with different temperatures (c-f). Note that in each case the temperature is  assumed to be uniform over the CME. The filter intensity ratio, however, changes with position, showing higher values at the leg parts of the CME.
 We conjecture that the non-uniform filter intensity ratio results from the different amount of the Doppler shift due to the asymmetric density strucuture and height effect. When the scattered light comes from a coronal electron to the observer, the LOS distance between the coronal electron and the solar limb plane determines the amount of the Doppler shift. Therefore, an asymmetric density structure can generate a non-uniform filter intensity ratio. Height is also related with the Doppler shift. Coronal electrons located at lower heights are less affected by Doppler shift because lots of photospheric radiation has large incidence angle against the radial motion of the CME. Despite this non-uniform filter intensity ratio, Figure \ref{fig:jkasfig3} clearly indicates that the average filter intensity ratio decreases with the  CME temperature.

 We focus on the top part of the CME structure (the blue cross position in Figure \ref{fig:jkasfig3}b), which gives the most reliable result because \citet{Thernisien2006} used this position to extract the density parameters ($n_{e}$, $\sigma_{t}$, and $\sigma_{l}$) of the CME. Figure \ref{fig:jkasfig4}a shows the simulated spectra for each CME temperature at this position, together with the filter response functions. As \citet{Cram1976} explained, it turns out that higher CME temperature produces a smoother spectrum. Additionally, we find that the two filters are located at the wavelengths where intensity variation is sensitive to temperature variation.

 Figure \ref{fig:jkasfig4}b shows the relationship between  CME temperature and filter intensity ratio. The filter intensity ratio varies from 1.43 to 1.24 as temperature increases from 0.5 to 2.0$\times 10^{6}$ K.  This anti-correlation is consistent with our expectation. The filter intensity reatio variation is large enough to measure the CME temperature from the observed data, according to \citet{Reginald2009}. Even though this variation depends on the assumed CME model, we expect that the model dependence may be weak. This is supported by the fact that the derived variation of filter intensity ratio is found to be comparable to the value obtained by \citet{Reginald2000} for the spherically symmetric K-coronal density distribution. Thus, we expect that the CME temperature is measurable from the filter intensity ratio for a veriety of CME models.

\section{Discussion\label{sec:diss}}

 The method of determining CME temperature based on the filter intensity ratio  has several advantages over other methods. The biggest one is that it allows temperature determination over a wide field of view, unlike previous spectroscopic observations that had to be confined to a finite number of slit locations. In addition, the observing setup is much simpler than other instruments and can be easily implemented in a coronagraph instrument.  It requires only 2 filter images, which is easier to produce than spectrograms or multi-filter images.

 In real observations, securing high signal to noise (S/N) is crucial to accurately measure  CME temperatures. The filter observations needed for this method have disadvantages compared to current white light coronagraphic observations with respect to the achievable S/N. The filter width (30 \AA) is quite narrower than the spectral window of the existing white light coronagraphs \citep{Brueckner1995, Howard2008}, leading to lower S/N for given exposure time. The problem of low S/N  is further exasperated because the waveband for this observation is at near UV wavelengths where the solar irradiance is relatively low and the quantum efficiency of conventional CCDs is low. Thus the instrumental specifications should be considered carefully when acquiring data for the filter intensity ratio method.

 We plan to improve the filter intensity ratio method by implementing density distributions other than the GCS model. Because the GCS model considers only 9 free parameters, there are discrepancies between the model and the reality. We confirmed that the filter intensity ratio does not depend on the absolute value of the electron density ($n_{e}$) which is the most ambiguous parameter. Verification using other methods is also important. Comparison with results from previous methods or in situ measurements will be a great help in acquiring more reliable results. In the future, we will also investigate the Doppler effect of the CME motion on the filter intensity ratio.


\acknowledgments

This work was supported by the KASI research fund for Space Weather Forecast Center. LASCO CME catalog is generated and maintained at the CDAW Data Center by NASA and The Catholic University of America in cooperation with the Naval Research Laboratory. SOHO is a project of international cooperation between ESA and NASA.



\end{document}